\documentclass[a4paper,
               keeplastbox,   
               ]{jacow}
\usepackage{pdfpages,multirow,ragged2e} %
\usepackage{amsmath} %
\makeatletter%
	\ifboolexpr{bool{xetex}}
	 {\renewcommand{\Gin@extensions}{.pdf,%
	                    .png,.jpg,.bmp,.pict,.tif,.psd,.mac,.sga,.tga,.gif,%
	                    .eps,.ps,%
	                    }}{}
\makeatother

\ifboolexpr{bool{xetex} or bool{luatex}} 
 {}                                      
 {\usepackage[utf8]{inputenc}}           

\usepackage[USenglish]{babel}

\ifboolexpr{bool{jacowbiblatex}}%
 {%
  \addbibresource{jacow-test.bib}
  \addbibresource{biblatex-examples.bib}
 }{}
\begin{document}

\title{Definition of tolerances and corrector strengths for the orbit control of the High-Energy Booster ring of the future electron-positron collider\thanks{This work was supported by the European Union’s Horizon 2020 Research and Innovation programme under grant no. 951754 (FCCIS).}}

\author{B. Dalena\thanks{barbara.dalena@cea.fr}, T. Da Silva, A. Chancé, A. Ghribi\textsuperscript{1} \\
Paris-Saclay University and CEA Irfu, 91191 Gif-sur-Yvette, France \\
		\textsuperscript{1}also at  CNRS GANIL, 14118 Caen, France}
	
\maketitle
\begin{abstract}
After the discovery of the Higgs boson at the LHC, particle physics community is exploring and proposing next accelerators, to address the remaining open questions on the underlying mechanisms and constituents of the present universe. One of the studied possibilities is FCC (Future Circular Collider), a 100-km-long collider at CERN \cite{FCC-CDR}. The feasibility study of this future proposed accelerator implies the definition of tolerances on magnets imperfections and of the strategies of correction in order to guarantee the target performances of the High Energy Booster ring. The efficiency of the correction scheme, used to control the orbit, directly bounds the corrector needs and magnet tolerances. Analytic formulae give a first estimation of the average RMS values of the required dipole correctors’ strengths and of the allowed magnets misalignments and field quality along the entire ring. The distribution of the correctors along the ring is simulated,
in order to verify the quality of the residual orbit after the proposed correction strategy and to compare it with the analytical predictions. First specifications of the orbit correctors strength and tolerances for the alignment of the main elements of the ring are presented. The limits of the studied correction scheme and method are also discussed.
\end{abstract}

\section{INTRODUCTION}
The aim of the present study is to first, determine the tolerances for the misalignment of the elements of the High Energy Booster (HEB) of the lepton version of the Future Circular Collider (FCC-ee). Then, to estimate the correctors' strength to correct for closed orbit perturbations caused by machine errors.
The type of errors considered are: 
\begin{Itemize}
    \item random dipole field error and random dipole roll ;
    \item quadrupole alignment errors ;
    \item beam profile monitors (BPM) alignment and reading errors ;
    \item sextupole alignment errors.
\end{Itemize}
The expected Root Mean Square (RMS) values of the residual orbit and corrector strengths in presence of the considered errors have been calculated according to the following formulae~\cite{LHC43, LHC501}.
\begin{eqnarray}
   x_\text{rms} & = & \frac{\pi}{\sqrt{2}\sin{\pi Q_x}}\frac{\Bar{\beta}}{\sqrt{N_{d}}} \left(\frac{\Delta B}{B}\right)_\text{rms} + \nonumber\\
           &  & \frac{\sqrt{N_{q}}}{\sqrt{2}\sin{\pi Q_x}\cos{\mu/2}}(\Delta q_{x})_\text{rms} +\nonumber\\
           &  & \frac{\sqrt{1/2}}{\left[ 1 + \sin(\mu/2) \right]}(\Delta\sigma_{x})_\text{rms}\label{eq:orbit_x}\\
   y_\text{rms} & = & \frac{\pi}{\sqrt{2}\sin{\pi Q_y}}\frac{\Bar{\beta}}{\sqrt{N_{d}}} (\Delta \theta)_\text{rms} + \nonumber\\
           &   & \frac{\sqrt{N_{q}}}{\sqrt{2}\sin{\pi Q_y}\cos{\mu/2}}(\Delta q_{y})_\text{rms} +\nonumber\\
           &   & \frac{\sqrt{1/2}}{\left[ 1 + \sin(\mu/2) \right]}(\Delta\sigma_{y})_\text{rms}\label{eq:orbit_y}
\end{eqnarray}

\begin{eqnarray}
(\delta_{x})_\text{rms} =& \biggl\{ \frac{\Bar{\beta}}{\beta_\text{max}} \left( n \left[ \frac{2\pi}{N_{d}} \left( \frac{\Delta B}{B} \right)_\text{rms} \right]^{2} + 
       2 (\Delta q_{x})_\text{rms}^{2} K_{q}^{2} L_{q}^{2} \right) + \nonumber\\
     &    \frac{1+2\cos^{2}(\mu)}{ 2(L_\text{cell}/2)^{2}\left[1+\sin(\mu/2)\right]^{2}}(\Delta\sigma_{x})_\text{rms}^{2} \biggr\}^{1/2}\\
(\delta_{y})_\text{rms} =& \biggl\{ \frac{\Bar{\beta}}{\beta_\text{max}} \left( n \left[ \frac{L_{d}}{\rho} \left( \Delta \theta \right)_\text{rms} \right]^{2} + 
  2 (\Delta q_{y})_\text{rms}^{2} K_{q}^{2} L_{q}^{2}\right)+ \nonumber\\
 & \frac{1+2\cos^{2}(\mu)}{2(L_\text{cell}/2)^{2}\left[1+\sin(\mu/2)\right]^{2}}(\Delta\sigma_\text{y})_\text{rms}^{2} \biggr\}^{1/2} 
\end{eqnarray}\label{eq:corr}
where $N_d$ is the number of dipoles, $Q_{x,y}$ the horizontal/vertical tunes, $\Bar{\beta}=\frac{L_\text{cell}}{\sin(\mu)}$ the mean betatron function, $L_\text{cell}$ the cell length, $\mu$ the phase advance per cell, and $N_q$ the number of quadrupoles.
For each of 100 different error configurations of the lattice model we compute the corresponding analytical value. We compare the maximum among the analytical RMS estimates with the distribution of 100 numerical results, after applying the correction strategy described in the next section.

\section{Correction Strategy}
\begin{figure}[htb]%
    \centering
    \includegraphics[width=0.33\textwidth]{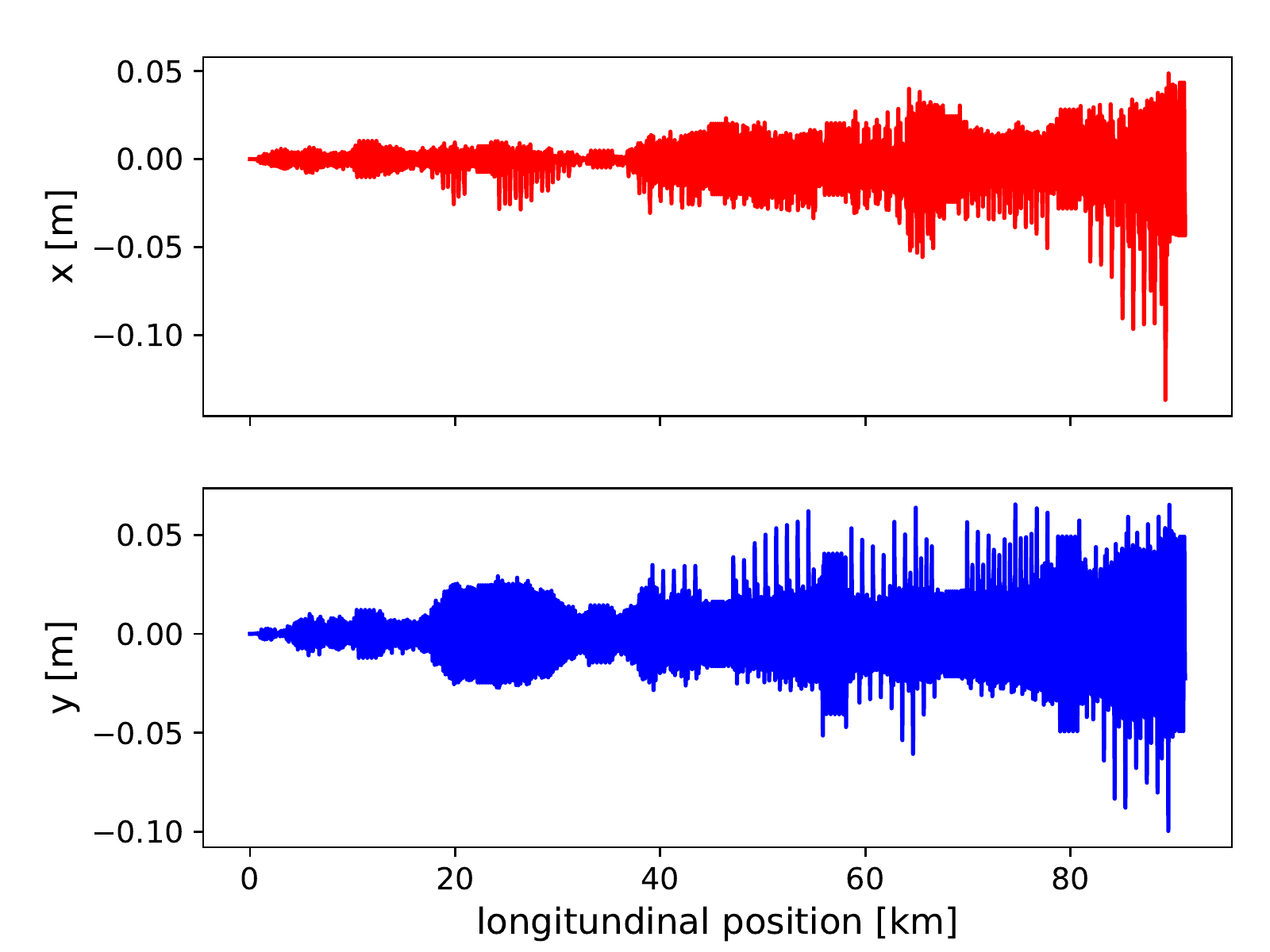}
    \includegraphics[width=0.33\textwidth]{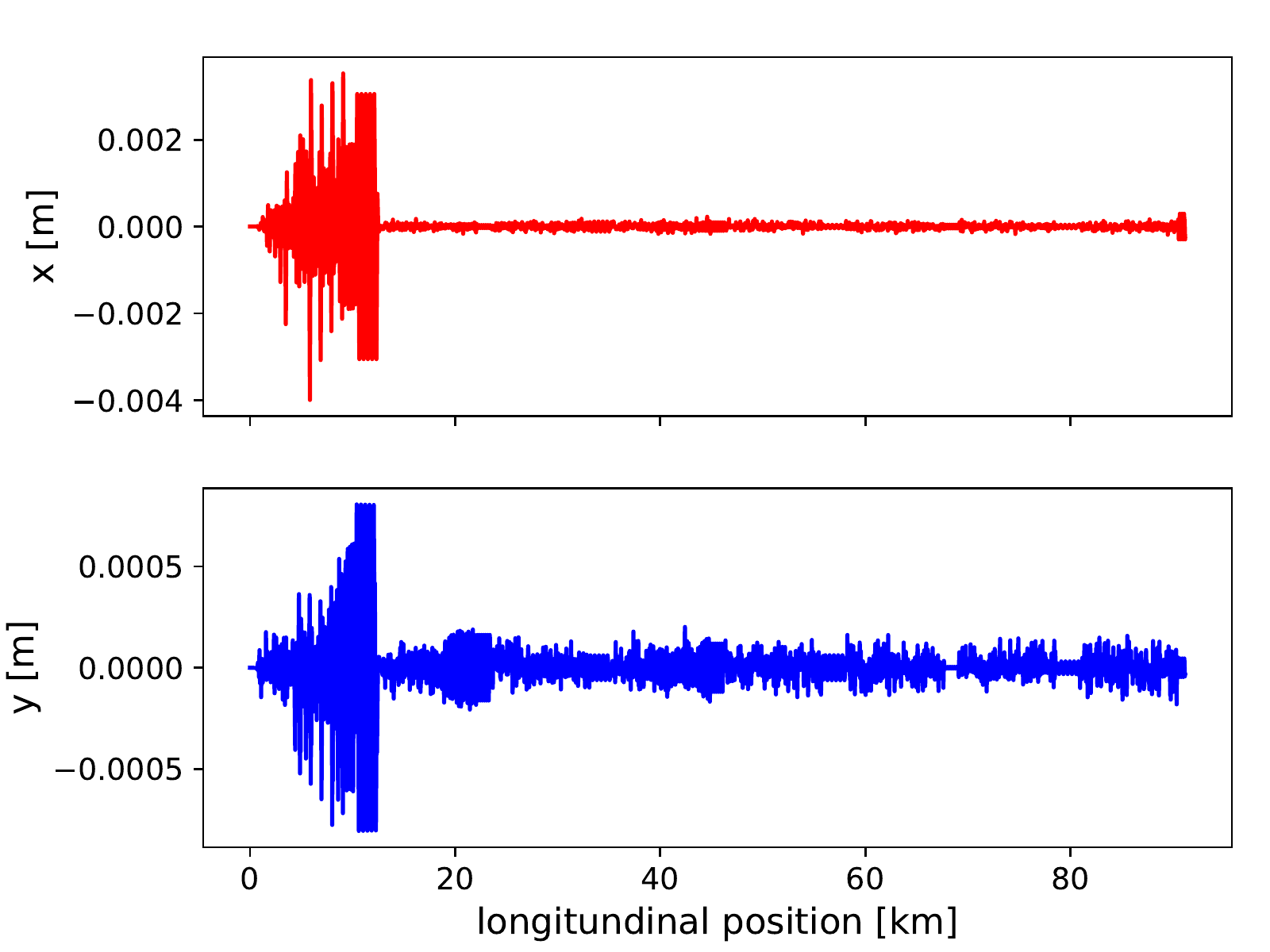}
    \includegraphics[width=0.33\textwidth]{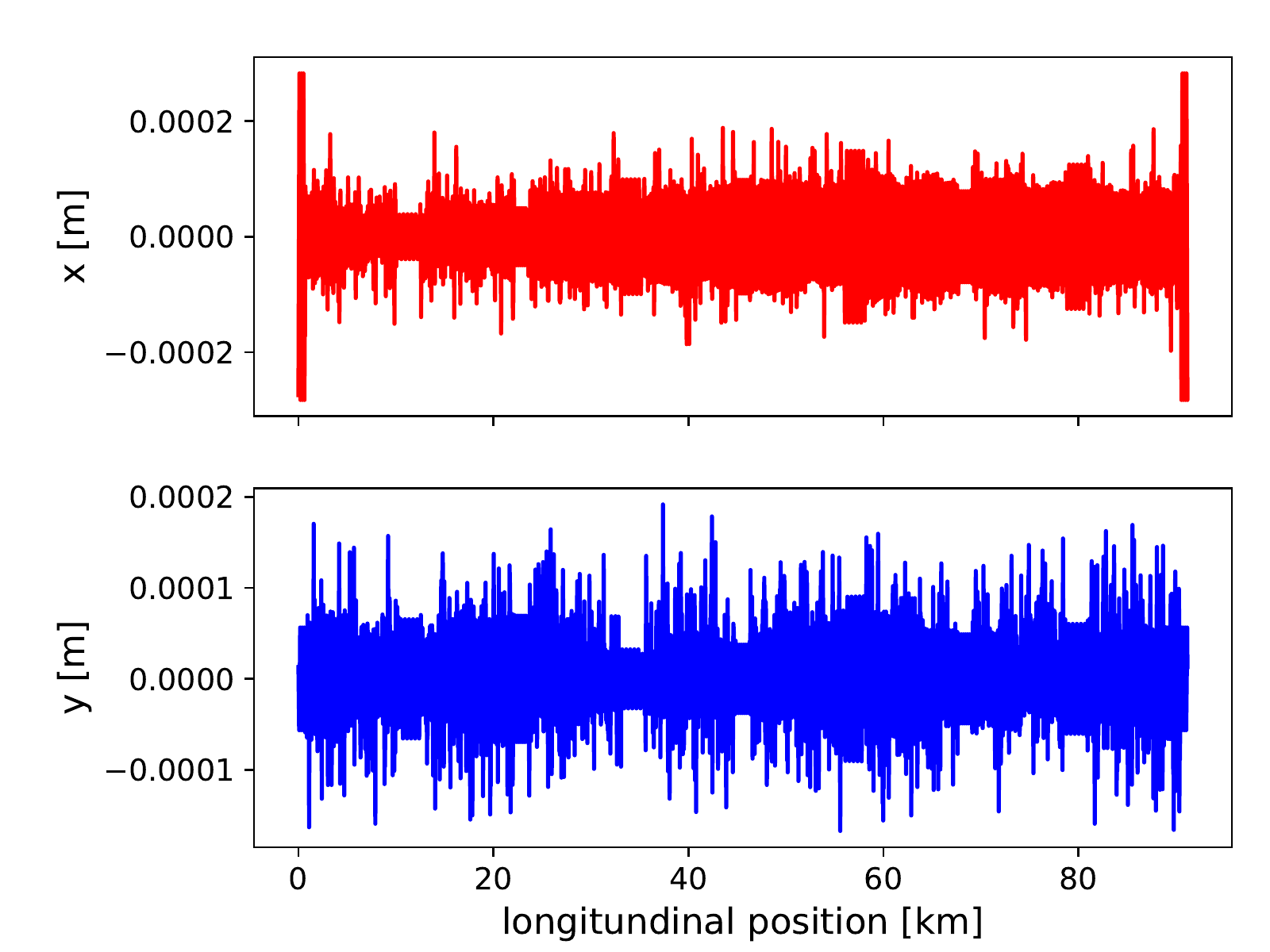}
    \caption{ Example of the evolution of the orbit during three different steps of the correction procedure (seed=3, MQ offset (\SI{150}{\micro\metre}), MB field error ($10^{-4}$), main MB roll (\SI{300}{\milli\radian}), BPM offset (\SI{150}{\micro\metre}), and MB offset (\SI{150}{\micro\metre})). Top panel shows residual orbit with errors and no correction. Middle panel shows residual orbit after last iteration of the SVD correction with sextupoles OFF. Bottom panel shows residual orbit after the SVD correction with sextupoles ON.}%
    \label{fig:seed3}
\end{figure}

Before correcting the orbit, we first need to define and choose the correctors and BPMs to use. If a quadrupole is focussing in the horizontal plane the neighbouring BPM will read in horizontal and the following corrector will correct in horizontal. Reciprocally, if the quadrupole is focussing in the vertical plane, the BPM and corrector will act in vertical. The range of correctors and BPMs selected for each arc is defined in order to have the same number of correctors and BPMs used in each plane. In the $x$ plane: from the first BPM and the two correctors before the beginning of the arc to the three BPMs and correctors after the end of the arc. In the $y$ plane: from the first BPM and the two correctors before the beginning of the arc to the two BPMs and correctors after the end of the arc.\\
The same strategy has been used for the LHC during commissioning~\cite{cern2}: a first correction, Segment-by-Segment (SbS) \textit{i.e.} in our case, arc by arc.
Each segment of the machine is handled as a line, using the optical parameters at the first segment entry. 
The correction procedure is divided in two sections: one with the sextupoles turned off and the other with the sextupoles on.
This decomposition sextupoles off/on of the orbit study is commonly done in accelerators such as SuperKEKB during commissioning \cite{SuperKEKB}.
We assume that all the correctors of the Booster will be individually powered. \\
\textbf{Procedure part 1: sextupoles off.} After turning off the sextupoles, we first do the SbS. This gives us a first orbit corrected to inject in a singular value decomposition (SVD) algorithm over the entire machine. The SVD is done with the MadX command \textit{CORRECT}~\cite{madx}. By doing an SbS before a SVD on all arcs, we can make more seeds converge than starting directly with all arcs together.
After the SbS, two iterations of SVD are made on all arcs and in line, in order to reduce the residual orbit. 
This is sufficient to get an orbit reduced enough to find the closed orbit. 
Finally, multiple iterations of SVD in ring are made and the number of iterations vary for each seed. Iterations are made until the RMS of the orbit in each plane is below the analytical RMS, while the number of iterations is under a limit of maximum iterations (fixed to 15).\\
\textbf{Procedure part 2: sextupoles on.} The sextupoles are turned on and we do one iteration of SVD in ring. More iterations are not useful because the command \textit{CORRECT} does not give a better correction of the residual orbit.\\
To illustrate the different steps of the correction procedure, the Fig.~\ref{fig:seed3} represents the orbit for the seed 3, with an RMS random error on the quadrupole offset of \SI{150}{\micro\metre}, dipole relative field error of $10^{-4}$, main dipole roll of \SI{300}{\milli\radian}, BPM offset of \SI{150}{\micro\metre} and sextupole offset of \SI{150}{\micro\metre}, after each important step.
We studied and tested different case scenarios by adding one by one error types. All tests were run on 100 seeds.
Our starting point is the combination of the offsets of the quadrupoles (designated as MQ), the dipole (designated as MB) relative field error and main dipole roll error. The statistical study
of this first error configuration revealed that all seeds converged until we reached a MQ offset of \SI{150}{\micro\metre}. By following this method, we added and fixed the other elements errors as reported in Table~\ref{tab:errtable}. 

\begin{table}[!htb]
\caption{Summary of the different error types and their values.}
\label{tab:errtable}
\small
\begin{tabular}{ll}
\hline
Error type  (Gaussian RMS ) & Value [Unit]                                       \\ \hline
MQ offset                 & 80, 90, 100, 120, 150, 200 \si{\micro\meter}   \\
MB relative field error   & 10$^{-4}$,10$^{-3}$      \\
MB main dipole roll error & \SI{300}{\milli\radian}\\
BPM offset                & 60, 80, 100, 150, 200 \si{\micro\metre} \\   \textit{MQ offset  \SI{150}{\micro\metre}} &  \\
MS offset                 & 60, 80, 100, 120, 150, 200 \si{\micro\metre} \\  \textit{MQ, BPM offset  \SI{150}{\micro\metre}}  & \\
BPM resolution            & 10 and 50 \si{\micro\metre}  \\                    \textit{MQ, BPM, MS offset \SI{150}{\micro\metre}} &    \\ 
\hline
\end{tabular}
\end{table}
It is worth noticing that all the errors applied on the elements are randomly Gaussian distributed within $\pm$3 RMS.

\section{Results and discussion}

\begin{figure}[th!]%
    \centering
    \includegraphics[width=0.35\textwidth]{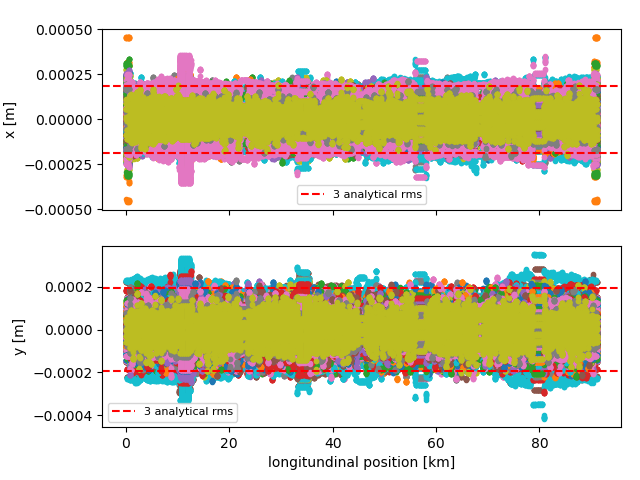}
    \includegraphics[width=0.35\textwidth]{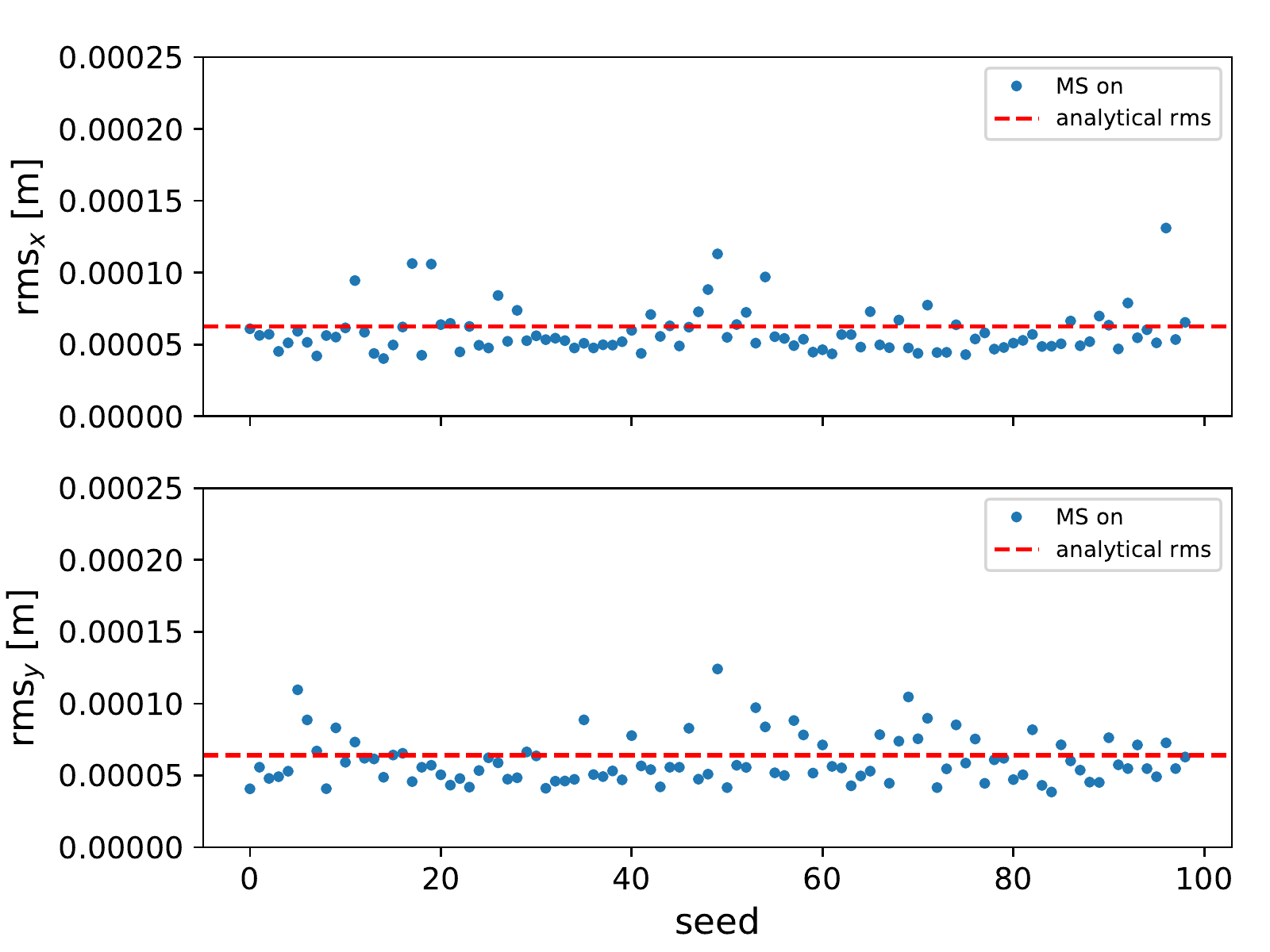}
    \caption{Distribution of the residual orbits and the corresponding RMS values for the 99 successful machine configurations and errors described in the text.}%
    \label{fig:orb_dist}%
\end{figure}

\begin{figure}[th!]
    \centering
    \includegraphics[width=0.35\textwidth]{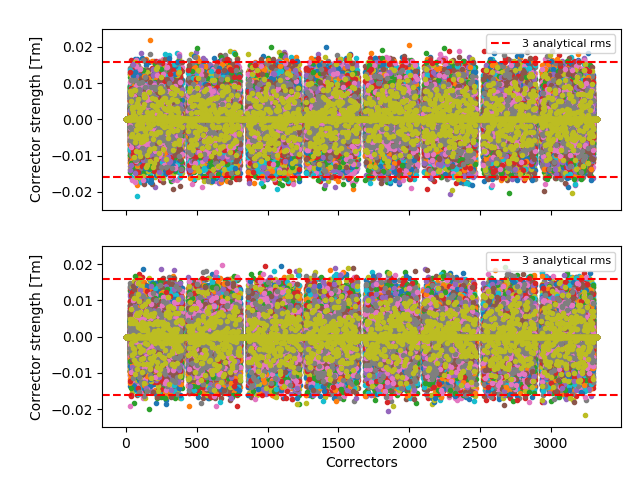}
    \includegraphics[width=0.35\textwidth]{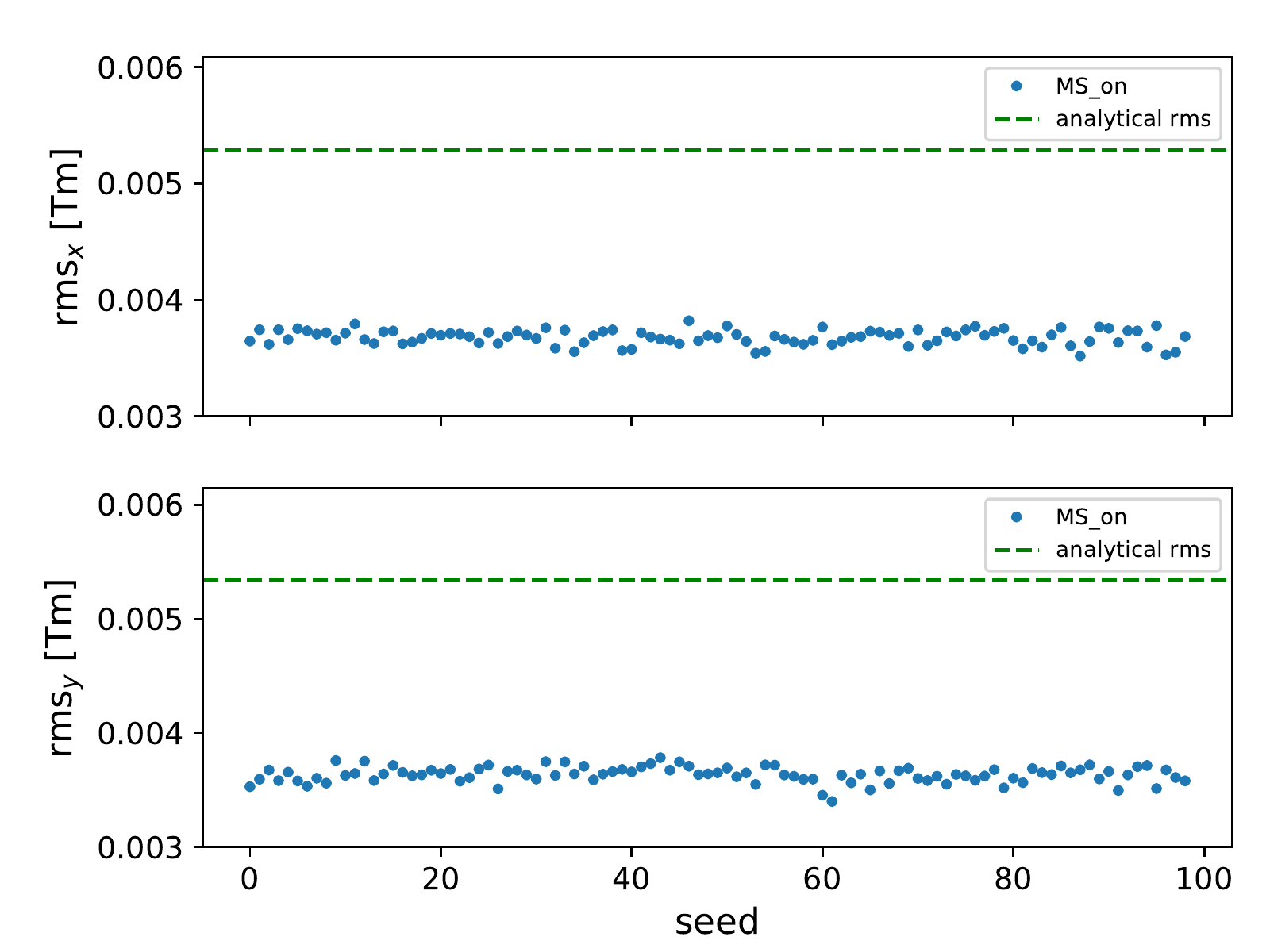}
    \caption{Distribution of the correctors strength and the corresponding RMS values for the 99 successful seeds.}%
    \label{fig:corr_dist}%
\end{figure}

Figures~\ref{fig:orb_dist} and~\ref{fig:corr_dist} show the orbit and correctors strength distribution with their respective RMS values for the 99 successful seeds at the end of the correction procedure.  The following errors are included: relative dipole field error of 10$^{-3}$, main dipole roll of \SI{300}{\milli\radian}, BPMs offset of \SI{150}{\micro\metre}, sextupole offset of \SI{150}{\micro\metre} and BPM reading error of \SI{50}{\micro\metre}. The dashed red lines on the distributions represents $\pm$3 times the RMS calculated analytically.
For both the orbit and the correctors' strength, we can see that our analytical predictions fit well the data from the simulations, for most all of the seeds. 
The deviations can be explained by the combination of the different errors and $\beta$-function, for which the procedure is less effective.
The residual orbit amplitude in both planes is in the order of magnitude of the MQ offsets (which are the dominant errors as expected) and the pattern of the succession of the arcs and insertions is visible (because we only applied the errors to the arcs, the residual orbit after correction in the insertions is expected to be almost zero). 
The RMS values for each of the successful seeds are distributed around the dashed red line representing one analytical RMS estimate, the blue dots correspond to the RMS after turning on the sextupoles and correcting the orbit once, always using the SVD algorithm.  
Apart from the QP offset, the other main contributors to the residual orbit are the MS offset and the BPM resolution. 
Concerning the correctors' strength distribution (Fig.~\ref{fig:corr_dist}), we can notice that for almost all the correctors, their strength is within the limit of the three analytical RMS estimates (red dashed line). The RMS values for the 99 successful seeds are the same between the last iteration of the first part of the correction procedure (sextupoles off) and its end (see sec. Correction Strategy). The analytical RMS (green dashed line) is well above the numerical results from the simulations.
The values of the corresponding residual orbit and corrector specifications is reported in Table~\ref{tab:sumtab}.

\begin{table}
\caption{Residual Orbit and Corrector strengths for \SI{150}{\micro\metre} RMS Gaussian distributed random offset values of quadrupoles, sextupoles, and BPMs, \SI{50}{\micro\metre} RMS Gaussian distributed random precision of BPM reading, 10$^{-3}$ RMS Gaussian distributed random relative dipole magnetic field error and \SI{300}{\milli\radian} RMS Gaussian distributed roll of dipoles.}\label{tab:sumtab}
\begin{minipage}[b]{\columnwidth}
\small
\begin{tabular*}{\columnwidth}{@{\extracolsep{\fill} }l*{3}{c}r}
\toprule
& & Plane & \multicolumn{2}{c}{3$\times$RMS}\\
& & & Analytic & Seeds\footnote{Mean of RMS of all seeds.}\\
\midrule
Residual orbit & [\si{\micro\metre}] & x & 188 & 174\\
& & y & 192 & 180\\
\midrule
Corrector strength & [\si{m\tesla$\cdot$\metre}] & x & 16 & 12\\
& & y & 16 & 12\\
\bottomrule
\end{tabular*}
\end{minipage}
\end{table}
\section{Conclusions}
We have computed the first specification for the tolerance of mis-alignment of the main high energy booster elements (of about \SI{150}{\micro\metre}) and dipole corrector strengths (of about \SI{20}{\milli\tesla$\cdot$\metre}), considering only the orbit corrections. These specifications do not consider the additional strength that will be required at the extraction energy to compensate the orbit deviation due to the energy loss in one turn because of synchrotron radiation (tapering). Moreover, these values need to be confirmed by the full emittance tuning (i.e. adding the $\beta$-beating, the dispersion and the coupling correction).
Finally, a further optimization of these specifications and of the correction strategy can be attempted, for example by mean of machine learning techniques. 

\section{ACKNOWLEDGEMENTS}
The Authors would like to thanks I. Agapov, R. Tomas and M. Hostettler for useful discussions.

%
%
\ifboolexpr{bool{jacowbiblatex}}%
	{\printbibliography}%

\begin{thebibliography}{9} 
        \bibitem{FCC-CDR}
        Future Circular Collider Study. Volume 2: The Lepton Collider (FCC-ee) Conceptual Design Report, preprint edited by M. Benedikt et al. CERN accelerator reports, CERN-ACC-2018-0057, Geneva, December 2018. Published in Eur. Phys. J. ST.

        
	\bibitem{LHC43}
		LHC Project Note 43\\
    \textit{Concerning the Strength of Closed Orbit Correction Dipoles in the Arcs of LHC V4.2}, John F. Miles, 12/02/1996\\
    \url{https://cds.cern.ch/record/691874/files/project-note-43.pdf}
	
    \bibitem{LHC501}
    LHC Project Note 501\\
    \textit{Field Quality Specification for the LHC Main Dipole Magnets}, Stéphane Fartoukh and Oliver Brüning, 10/2001\\
    \url{https://cds.cern.ch/record/522049/files/lhc-project-report-501.pdf}

    \bibitem{SuperKEKB}
    Yukiyoshi Ohnishi \textit{et al.}, Accelerator design at SuperKEKB, Progress of Theoretical and Experimental Physics, Volume 2013, Issue 3, March 2013, 03A011, \url{https://doi.org/10.1093/ptep/pts083}

    \bibitem{cern2}
    Physical Review Special Topics - Accelerators and Beams\\
    \textit{First $\beta$-beating measurement and optics analysis for the CERN Large Hadron Collider}, M. Aiba, S. Fartoukh, A. Franchi, M. Giovannozzi, V. Kain, M. Lamont, R. Tomás, G. Vanbavinckhove, J. Wenninger and F. Zimmermann, 13/08/2009\\
    \url{https://journals.aps.org/prab/pdf/10.1103/PhysRevSTAB.12.081002}
     
     \bibitem{madx}
     MADX documentation\\
     \textit{The MAD-X Program, User’s Reference Manual},
      L. Deniau, H. Grote, G. Roy, F. Schmidt, 25/02/2022\\
	   \url{https://mad.web.cern.ch/mad/webguide/manual.html}
    
	\end{thebibliography}
	{%
	

} 
%
%


\end{document}